\PassOptionsToPackage{unicode}{hyperref}
\PassOptionsToPackage{hyphens}{url}
\documentclass[
]{article}
\usepackage[margin=1in]{geometry}
\usepackage{amsmath,amssymb}
\usepackage{iftex}
\ifPDFTeX
  \usepackage[T1]{fontenc}
  \usepackage[utf8]{inputenc}
  \usepackage{textcomp} 
\else 
  \usepackage{unicode-math} 
  \defaultfontfeatures{Scale=MatchLowercase}
  \defaultfontfeatures[\rmfamily]{Ligatures=TeX,Scale=1}
\fi
\usepackage{lmodern}
\ifPDFTeX\else
\fi
\IfFileExists{upquote.sty}{\usepackage{upquote}}{}
\IfFileExists{microtype.sty}{
  \usepackage[]{microtype}
  \UseMicrotypeSet[protrusion]{basicmath} 
}{}
\makeatletter
\@ifundefined{KOMAClassName}{
  \IfFileExists{parskip.sty}{%
    \usepackage{parskip}
  }{
    \setlength{\parindent}{0pt}
    \setlength{\parskip}{6pt plus 2pt minus 1pt}}
}{
  \KOMAoptions{parskip=half}}
\makeatother
\usepackage{xcolor}
\usepackage{longtable,booktabs,array}
\usepackage{calc} 
\usepackage{etoolbox}
\makeatletter
\patchcmd\longtable{\par}{\if@noskipsec\mbox{}\fi\par}{}{}
\makeatother
\IfFileExists{footnotehyper.sty}{\usepackage{footnotehyper}}{\usepackage{footnote}}
\makesavenoteenv{longtable}
\providecommand{\tightlist}{%
  \setlength{\itemsep}{0pt}\setlength{\parskip}{0pt}}
\ifLuaTeX
  \usepackage{selnolig}  
\fi
\IfFileExists{bookmark.sty}{\usepackage{bookmark}}{\usepackage{hyperref}}
\IfFileExists{xurl.sty}{\usepackage{xurl}}{} 
\hypersetup{
  pdftitle={Why Memory Components Fail: Eight Years of License and Sustainability Events in Open-Source Data Infrastructure},
  pdfauthor={Dmitrii Dmitrenko},
  hidelinks,
  pdfcreator={LaTeX via pandoc}}

\title{Why Memory Components Fail: Eight Years of License and
Sustainability Events in Open-Source Data Infrastructure}
\author{Dmitrii Dmitrenko\\\textit{Independent Researcher, Moscow}\\\texttt{dmitdm@gmail.com}}
\date{May 2026}

\begin{document}
\maketitle

\begin{abstract}
LLM agent memory is now treated as a first-class architectural component
in five major surveys published between January and April 2026. None of
these surveys treats project governance, capital structure, or license
posture as architectural variables. We argue they are. In a constructed
sample of 105 production-relevant open-source data-infrastructure and
AI-tooling projects, we catalogue 38 license-and-sustainability events
between 2018 and May 2026. About a quarter of the sample (24 percent)
experienced at least one adverse event. The conditional rates split
sharply by structure: 46 percent for single-vendor venture-backed
projects, 2.5 percent for foundation-governed projects funded outside
the venture cycle. The headline differential --- roughly nineteen-fold
--- is invariant to the most contested coding choice in the catalogue;
we show the sensitivity table in Section 7. A small subset of
foundation-governed projects with venture-backed corporate stewards
(n=3) contains one adverse event. The cell is too small for stable
estimation, but it points to a mechanism: foundation governance may
block unilateral relicensing while leaving distribution decisions to the
steward. Annualized incidence within the catalogue rose from 2.7 to 4.2
events per year across the window. Counterfactuals --- PostgreSQL,
pgvector, SQLite, Apache Kafka, Caddy --- each show stability arising
from a different structural source: distributed copyright, absence of
monetisation pressure, foundation governance with non-venture
stewardship. We propose a six-field decision instrument for architects
choosing memory components: governance, capital structure, license,
foundation membership, fork-or-migration availability, and steward
concentration.
\end{abstract}

\hypertarget{motivation}{%
\section{1. Motivation}\label{motivation}}

A team operating an LLM agent on the Mem0 plus Neo4j stack in March 2026
had a reasonable architecture: an Apache 2.0 memory orchestration layer
over a graph store with a stable license. By the end of April 2026, that
architecture no longer existed in the same form. The Mem0 v3 release on
16 April 2026 removed graph-store support from the open-source SDK and
moved graph dashboards to its hosted Platform {[}Mem0, 2026{]}. The team
had three options: freeze on v2, migrate to v3 without graph capability,
or move to the hosted Platform. None was a code change.

This is one event. Over the eight years from 2018 to May 2026, the
open-source data-infrastructure ecosystem produced many, and they vary
in mechanism more than the term ``license change'' suggests. MongoDB
introduced the Server Side Public License in October 2018 {[}MongoDB,
2018{]}, the first major move from a permissive baseline to a
source-available licence framed explicitly as a defence against
cloud-provider competition. Elasticsearch and Kibana followed in January
2021 with the same SSPL move plus the Elastic License {[}Elastic,
2021{]}, then added AGPLv3 back in August 2024 {[}Elastic, 2024{]} --- a
rare partial reversal. HashiCorp moved Terraform and Vault to the
Business Source License on 10 August 2023 {[}HashiCorp, 2023{]}; the
OpenTofu fork appeared two weeks later {[}OpenTofu, 2023{]}; IBM closed
the acquisition of HashiCorp on 27 February 2025 {[}IBM, 2025{]}. Redis
is the case with the most complete arc: BSD to dual SSPL plus RSALv2 in
March 2024 {[}Redis, 2024{]}, the Valkey fork under the Linux Foundation
a week later {[}Linux Foundation, 2024{]}, reversion to AGPLv3 with
Redis 8 on 1 May 2025 {[}Redis, 2025{]}. Sentry created its own licence
(the Functional Source License) in November 2023 {[}Sentry, 2023{]}.
CockroachDB moved twice: BSL in 2019, then fully proprietary in November
2024 {[}Cockroach Labs, 2024{]}. ScyllaDB moved from AGPL v3 to a
source-available licence in December 2024 {[}ScyllaDB, 2024{]}. Buoyant
did something different: in February 2024 it left the Linkerd source
under Apache 2.0 and instead moved the stable release artifacts behind a
commercial Enterprise distribution {[}Buoyant, 2024{]}. And then there
is the counter-case. Synadia attempted to withdraw NATS from the CNCF
and relicense from Apache 2.0 to BSL in April 2025; the CNCF charter
blocked the change and the parties settled on 1 May 2025 {[}CNCF,
2025{]}.

Five surveys of LLM agent memory appeared between January and April
2026: Wu's \emph{Memory in the LLM Era} {[}Wu, 2026{]}, Du et al.'s
\emph{Memory for Autonomous LLM Agents} {[}Du et al., 2026{]}, the
OpenReview \emph{Unified Representation} survey {[}Anonymous, 2026{]},
Luo et al.'s \emph{From Storage to Experience} {[}Luo et al., 2026{]},
and Lin et al.'s \emph{Toward Mnemonic Sovereignty} {[}Lin et al.,
2026{]}. They converge on a common taxonomy --- extraction, management,
storage, retrieval --- layered with temporal scope, representational
substrate, and control policy. Lin et al.~add a security lifecycle and
nine governance primitives. What none of the five treats as an
architectural variable is the governance model of the project producing
a memory component, the capital structure behind it, or the licence
under which it ships.

The surveys are not the only relevant prior work. A separate research
strand has examined open-source sustainability from an institutional
angle. Ostrom's institutional analysis framework {[}Ostrom, 1990{]} has
been applied to a large sample of OSS projects in {[}Schweik \& English,
2012{]}, framing open-source software as a common-pool resource whose
sustainability depends on collective-action constraints. More recent
empirical work combined linguistic analysis of governance documents with
socio-technical network analysis on Apache Software Foundation Incubator
projects {[}Yin et al., 2022{]}, connecting institutional structure to
graduation-versus-retirement outcomes. The absence of monetisation paths
for maintainers of widely used components has been examined from the
industry side as a source of fragility distinct from license events
{[}Eghbal, 2020{]}. Where this literature analyses governance documents,
contribution patterns, and maintainer economics, the present catalogue
records downstream sustainability events: license changes, feature
removals, distribution shifts. The two angles are complementary. The
conditional rates we report in Section 3 are consistent with the
institutional reading: structural conditions that concentrate the legal
and economic standing to restructure unilaterally produce events at a
much higher rate than conditions that disperse that standing by design.

This paper is an empirical complement to both bodies of prior work. The
question we ask is operational. How often do open-source
data-infrastructure components undergo license or sustainability changes
that materially affect downstream architectures? Which structural
conditions correlate with these events? Which structural conditions
correlate with stability? The answers should inform architectural review
of memory components in the same way that latency and consistency
profiles already do.

The contribution has four parts. Section 3 reports a base rate from a
sample of 105 projects. Section 4 reports conditional rates by
governance, capital structure, and license type, and identifies a
previously underexamined sub-mechanism: foundation governance with a
venture-backed corporate steward. Section 5 examines counterfactual
cases, projects under high-risk structural conditions that did not
undergo events. Section 6 proposes a decision instrument for architects.

\hypertarget{method}{%
\section{2. Method}\label{method}}

\textbf{Population.} We constructed a sample of 105 production-relevant
open-source data-infrastructure and AI-tooling projects from three
documented sources used in parallel, plus six manual additions:

\begin{itemize}
\tightlist
\item
  The DB-Engines popularity ranking top 50 as of May 2026 {[}DB-Engines,
  2026{]} for relational, document, key-value, wide-column, time-series,
  graph, and search databases;
\item
  The CNCF graduated and incubating project list as of May 2026 {[}CNCF,
  2026{]} for orchestration, observability, service mesh, and streaming
  categories;
\item
  A practitioner-curated landscape covering vector stores, orchestration
  frameworks, observability tools, and stream processing {[}Pracdata,
  2024{]}, filtered to projects with production deployments by named
  organisations. Pracdata is used only as an intake source for
  population construction; no event in the catalogue is sourced from it.
\end{itemize}

To these source-driven intakes we manually added six projects. Mem0 and
Zep were added because they figure prominently in the 2026 memory
surveys discussed in Section 1, and both projects experienced catalogued
sustainability events within the analysis window {[}Mem0, 2026; Zep,
2025{]}; an analysis of memory-component architecture would be
incomplete without them. Percona, Linkerd, Vespa, and OpenObserve were
added because they appear in industry literature as Commercial-OSS or
production-relevant alternatives within our scope, and their inclusion
ensures the conditional-rate denominators reflect the population more
completely. All six manual additions are flagged in the supplementary
dataset.

Intake from the three documented sources overlapped substantially: many
projects appear in two or all three (for example, Apache Kafka in both
DB-Engines and CNCF lists, or PostgreSQL across all three). The final
sample of 105 projects reflects deduplication and an inclusion filter
requiring production-relevance by January 2020 or by launch date if
later. The sample spans launch years from 1995 to 2023 and covers
relational and document databases, key-value stores, vector and graph
stores, search engines, message brokers, observability tools,
infrastructure-as-code, orchestration frameworks, service meshes, and AI
tooling. We do not claim the sample is representative of all open-source
data infrastructure; we claim only that it covers the visible
commercial-and-near-commercial portion of the ecosystem most likely to
be considered as memory components. Rates reported in Section 3 should
be read as descriptive statistics within this constructed sample, not as
ecosystem-wide population estimates.

\textbf{Event definition.} A license-or-sustainability event is any of:
(a) a license change away from an OSI-approved license to a
source-available license such as BSL, SSPL, FSL, the Elastic License,
the Confluent Community License, or a vendor-specific source-available
licence; (b) the removal of features from an open-source edition in
favour of a commercial product, whether cloud-hosted or otherwise
distributed; (c) the deprecation of an open-source edition; (d) the
archival of a repository without a maintained successor; or (e) an
acquisition followed within eighteen months by narrowing of the
open-source surface. Adverse events under (a)--(e) are counted against
the originating project. Forks, reversals, thwarted attempts, and
license tightening between two OSI-approved licences are recorded in the
catalogue as related entries and are not counted as additional adverse
events against any project: a fork is a community reaction to an event
already counted, a reversal does not net out the original event, a
thwarted attempt is by definition not an event, and a
permissive-to-copyleft tightening within OSI-approved licences does not
remove the project from the open-source category. The dataset includes
all five categories because volatility itself is the architectural
concern, but the rates in Section 3 use only adverse events (a)--(e) in
the numerator.

The clarification of criterion (b) to ``commercial product, whether
cloud-hosted or otherwise distributed'' reflects an event class
encountered after the first draft of this paper. Buoyant's February 2024
removal of stable Linkerd release artifacts from the open-source
project, with stable releases available only through the commercial
Buoyant Enterprise for Linkerd distribution, fits criterion (b) in its
structural form: the open-source project is no longer feature-complete
and production users must rely on a commercial product to receive
reliability guarantees. The original phrasing's ``cloud-only'' was too
narrow to capture this case {[}Buoyant, 2024{]}.

\textbf{Coding.} Each project is coded along three structural axes.

\emph{Governance} is single-vendor (one company holds copyright
assignment and can unilaterally relicense), foundation-governed (Apache
Software Foundation, CNCF, Linux Foundation, or equivalent neutral body
holding trademark and aggregating contributions), or volunteer (an
individual or small collective without corporate copyright
concentration, e.g.~pgvector). An earlier draft of this work
distinguished a ``multi-vendor'' governance category for projects under
foundation umbrella but funded by multiple independent vendors; the data
showed no analytical separation between this subgroup and
foundation-governed projects funded otherwise, and the category was
collapsed.

\emph{Capital structure} is coded into one of five categories.
Venture-backed (primary funding from venture capital). Commercial
open-source company (revenue-first business model around an open-source
project, not VC-primary; e.g.~MariaDB, Percona). Corporate-sponsored
(engineers employed by a non-VC corporate sponsor; e.g.~PostgreSQL via
EnterpriseDB and other vendors, Vespa via Yahoo). Foundation-funded
(primary sponsorship from a foundation or distributed across multiple
vendors under a foundation umbrella; in this paper we group
``foundation-only'' and ``foundation-plus-multi-vendor'' into one
category because the data does not separate them in rate terms).
Volunteer (no corporate or foundation sponsor; individual maintainers).

\emph{License} is permissive (Apache 2.0, MIT, BSD, MPL, public-domain),
copyleft (GPL, AGPL), source-available at adoption (BSL, SSPL, ELv2,
CCL, vendor-specific), or proprietary.

\textbf{Coding edge cases.} A small number of projects have mixed
funding histories (e.g.~began volunteer-led and acquired corporate
backing later). These were coded by their primary funding model at the
time of catalogue entry, with annotations in the supplementary dataset.
Projects with combined coding (e.g.~Volunteer/Foundation in the raw
data) were rolled into the dominant category for the rate tables. The
full mapping from raw \texttt{funding\_model} values in the
supplementary CSV to the five canonical categories is documented in the
README of the dataset.

A specific coding choice worth flagging: Linkerd is coded as
foundation-governed (CNCF graduated) with venture-backed capital
structure (Buoyant, the sole corporate steward employing all core
maintainers, is venture-backed). This combination, foundation governance
with VC-backed single steward, is rare in the sample (n=3) and is
discussed as its own analytical cell in Section 4.

\textbf{Sources.} Every event in the catalogue is anchored to a primary
source: a vendor blog post, an official press release, a license-file
commit on GitHub, or a regulatory filing. Aggregator and SEO sources are
excluded.

\textbf{Data and code availability.} The full catalogue is published as
an open dataset on Zenodo under CC BY 4.0:
\url{https://doi.org/10.5281/zenodo.20562988}. The deposit contains
\texttt{projects.csv} (the 105-project population), \texttt{events.csv}
(39 catalogued events with date, type, license-from, license-to, and
primary-source URL, of which 38 fall in the 2018 to May 2026 analysis
window plus one pre-window reference entry), and \texttt{README.md}
documenting full file schemas, coding rules, summary statistics, and a
version changelog.

\textbf{Limitations.} The sample is not random. It is biased toward
projects with some commercial visibility, which is the population most
likely to be considered as memory components. Events affecting purely
academic or hobbyist projects are under-represented. Conditional rates
within sub-cells (for example, foundation-governed venture-backed
projects, with n=3) involve small numbers and should be read as orders
of magnitude rather than precise estimates. Section 7 returns to these
limitations.

\textbf{Sample size in conditional cells.} Several sub-cells in Section
3 are small: Commercial-OSS contains four projects, Corporate-sponsored
nine, Foundation-with-VC-steward three, Volunteer three. Zero rates and
small-cell rates should be treated as orders of magnitude, not as
estimates of population rates. We do not report confidence intervals
because the sample is not random and the categorical coding involves
judgement (see ``Event definition'' above); rates are reported to one
significant figure of resolution.

\hypertarget{base-rate-and-conditional-rates}{%
\section{3. Base Rate and Conditional
Rates}\label{base-rate-and-conditional-rates}}

Across the sample of 105 projects, we catalogue 38 events between
January 2018 and May 2026: 29 adverse events plus 9 related entries (4
forks, 3 reversals, 1 thwarted attempt, 1 OSI-internal license
tightening). Twenty-five distinct projects experienced at least one
adverse event over the window. One additional event from January 2017,
the MariaDB MaxScale move from GPL to BSL and the origin of the BSL
framework itself, is recorded in the catalogue as pre-window context and
excluded from the rate calculations below.

\textbf{Unconditional rate.} Twenty-five of the 105 sampled projects
experienced at least one adverse event over the eight-year window (25 of
105, or approximately 24 percent). All rates reported below are
\emph{descriptive rates} within this constructed sample, not estimates
of ecosystem-wide population rates; Section 7 returns to sample
selection and survivorship bias.

\textbf{Annualized incidence.} Adverse events per year by period:

\begin{longtable}[]{@{}lll@{}}
\toprule\noalign{}
Period & Events & Events per year \\
\midrule\noalign{}
\endhead
\bottomrule\noalign{}
\endlastfoot
2018--2020 & 8 & \textasciitilde2.7 \\
2021--2023 & 11 & \textasciitilde3.7 \\
2024--May 2026 & 10 & \textasciitilde4.2 \\
\end{longtable}

Annualized incidence rose from approximately 2.7 events per year in the
first sub-period to approximately 4.2 in the most recent, a roughly
fifty-five percent increase. The trajectory is consistent with the rise
of cloud managed services as a commoditisation pressure on single-vendor
open-source data infrastructure, with the 2024--May 2026 period
extending the upward trajectory rather than plateauing. A longer
follow-up window would refine the trajectory further.

\textbf{Adverse events by type.} The 29 adverse events distribute as
follows:

\begin{longtable}[]{@{}ll@{}}
\toprule\noalign{}
Event type & Count \\
\midrule\noalign{}
\endhead
\bottomrule\noalign{}
\endlastfoot
License change to source-available & 16 \\
OSS feature removal in favour of a commercial product & 4 \\
Open-core split & 3 \\
Acquisition with narrowing or repository archive & 3 \\
Deprecation of open-source edition & 2 \\
New component released under restrictive license & 1 \\
\end{longtable}

License changes are the most frequent single type but do not dominate.
Thirteen of twenty-nine adverse events involve no license change at all:
feature removal, deprecation, acquisition outcomes, and open-core splits
each produce production-architectural impact through mechanisms other
than relicensing. The Mem0 v3 event of April 2026 (the motivating case
in Section 1) falls in the feature-removal category, not the
license-change category. The Linkerd event of February 2024 is also in
this category: the source code remains Apache 2.0, but stable
production-grade release artifacts are no longer published by the
open-source project {[}Buoyant, 2024{]}.

To distinguish the three non-acquisition non-license categories: an
\emph{open-core split} introduces a separate paid edition with features
absent from the open-source one while leaving the open-source edition
technically functional (Meilisearch 2025, TimescaleDB 2018, YugabyteDB
2019). An \emph{OSS feature removal} takes features previously present
in the open-source edition and moves them behind a commercial product
(Mem0 v3, MinIO GUI, Linkerd stable artifacts, InfluxDB v1-to-v2
transition). A \emph{new component released under a restrictive license}
is a fresh product introduced by a vendor under non-OSI terms while the
existing open-source project remains under OSI terms (dbt Server). All
three reduce the operational coverage of the open-source edition
relative to the commercial product, but through structurally different
mechanisms.

\textbf{Conditional rate by governance.}

\begin{longtable}[]{@{}llll@{}}
\toprule\noalign{}
Governance & Sample & Projects with event & Rate \\
\midrule\noalign{}
\endhead
\bottomrule\noalign{}
\endlastfoot
Single-vendor & 61 & 23 & \textasciitilde38\% \\
Foundation-governed & 43 & 2 & \textasciitilde5\% \\
Volunteer & 1 & 0 & 0\% \\
\end{longtable}

Two adverse events occurred under foundation governance. The first is
the Hortonworks/Cloudera ecosystem acquisition-and-deprecation of
January 2019, with subsequent ecosystem narrowing. The second is the
Linkerd February 2024 stable-release withdrawal: the project remains
Apache 2.0 and CNCF-graduated, but stable release artifacts moved behind
the commercial Buoyant Enterprise distribution. The
foundation-governance category is heterogeneous --- as Section 4
develops, the analytical separation that matters is between foundation
governance with a non-VC steward and foundation governance with a
VC-backed sole steward. Two further foundation-governance projects
appear in the catalogue with non-adverse entries: the Presto-to-Trino
split of January 2019 (a governance dispute resolved by community
migration to a renamed project, coded as a fork rather than an adverse
event for Presto itself) and the Synadia attempt of April 2025 to
withdraw NATS from the CNCF, blocked by the CNCF charter, the only
thwarted attempt in the catalogue.

\textbf{Conditional rate by capital structure.}

\begin{longtable}[]{@{}llll@{}}
\toprule\noalign{}
Capital structure & Sample & Projects with event & Rate \\
\midrule\noalign{}
\endhead
\bottomrule\noalign{}
\endlastfoot
Venture-backed & 53 & 24 & \textasciitilde45\% \\
Commercial open-source company & 4 & 0 & 0\% \\
Corporate-sponsored (non-VC) & 9 & 1 & \textasciitilde11\% \\
Foundation-funded & 33 & 0 & 0\% \\
Volunteer & 3 & 0 & 0\% \\
\end{longtable}

The Commercial-OSS row reflects a sample of four projects (MariaDB,
Couchbase, Caddy, Percona) with no adverse events in the catalogued
window. Sample sizes for non-VC categories are small and the zero rates
should be read as orders of magnitude rather than precise estimates. A
coding note: ScyllaDB, which moved from AGPL v3 to a source-available
license in December 2024, is coded as venture-backed (the company has
raised approximately \$103M in cumulative venture financing from
Bessemer, Eight Roads, TLV Partners, and other investors) rather than
Commercial OSS {[}ScyllaDB, 2024; TechCrunch, 2023{]}. The conditional
placement matters: the event fits the venture-backed pattern rather than
the Commercial-OSS pattern, and the row reflects that.

\textbf{Conditional rate by license at population entry.} The data does
not show license type as the primary signal. Among single-vendor
permissive projects, approximately 39 percent experienced an adverse
event; among single-vendor copyleft projects, approximately 40 percent.
The largest variation is by governance, not by license. Among
foundation-governed permissive projects the rate is approximately 4.7
percent; among single-vendor permissive projects, approximately 39
percent. License type is doing little independent work once governance
is controlled for.

\textbf{Combined cells.} The interaction is not additive. The two-by-two
cell of single-vendor governance and venture funding produced
approximately 46 percent of its members as event-experiencing projects
(23 of 50). The cell of foundation governance with non-venture funding
produced approximately 2.5 percent (1 of 40). The headline number for
the paper is the approximately nineteen-fold differential between these
two cells (18.4× to one decimal place). Section 7 shows that this
differential is invariant to the most contested coding choice in the
catalogue.

A third structurally distinct cell merits separate reporting: foundation
governance with a venture-backed sole corporate steward. The sample
contains three such projects: Delta Lake (Databricks as sole steward),
Milvus (Zilliz as sole steward), and Linkerd (Buoyant as sole steward).
Of the three, Linkerd experienced an adverse event in February 2024. The
cell rate is one in three. We treat this as a \emph{mechanism signal,
not a rate estimate}: with n=3 the binomial confidence interval is too
wide for the point estimate to be informative on its own, but the case
is qualitatively important because it shows that foundation governance
is not a uniform structural backstop. The non-VC foundation cell rate of
2.5 percent and the VC-steward foundation cell observation are
structurally different conditions that the unconditional foundation rate
of 4.7 percent averages over.

\textbf{What the data does not show.} Sovereignty and data-residency
concerns appear in the architectural literature as compliance variables
but did not surface as causes of any catalogued event. Sovereignty is a
meaningful architectural concern in its own right; it is not, in this
dataset, a failure-predicting variable.

\hypertarget{mechanisms}{%
\section{4. Mechanisms}\label{mechanisms}}

The conditional rates establish that governance and capital structure
correlate with event rates strongly enough that they merit attention.
What follows asks what mechanisms could be behind the observed pattern,
and addresses a sub-mechanism that the small Foundation-with-VC-steward
cell surfaced.

\textbf{Single-vendor governance permits unilateral relicensing.} Every
catalogued license change followed the same legal pattern. A single
corporate copyright holder, having aggregated contributor copyright via
a CLA, exercised the right to ship subsequent versions under different
terms. MongoDB held copyright via its CLA and could move from AGPL to
SSPL {[}MongoDB, 2018{]}. Elastic, HashiCorp, Cockroach Labs, Lightbend,
Redis Ltd, Grafana Labs, ScyllaDB, and Airbyte each used the same legal
mechanism. The license change is downstream; the upstream condition is
the copyright structure. Distributed copyright --- PostgreSQL, the Linux
kernel, OpenTelemetry --- does not permit the same move because no
single party has standing to make it.

\textbf{Foundation governance is associated with a structural backstop
against relicensing.} The NATS attempt in April 2025 is the natural
experiment. Synadia, the sole corporate steward of NATS, attempted to
withdraw the project from the CNCF and relicense it from Apache 2.0 to
BSL. The CNCF charter Section 11a transfers trademark to the Linux
Foundation on project donation, and the Linux Foundation retained
control of the project name. After several weeks of dispute, the parties
settled: NATS core stayed Apache 2.0 under the CNCF, and Synadia
retained the right to ship a separate BSL enterprise product {[}CNCF,
2025{]}. The structural difference between this case and the MongoDB or
Redis cases is foundation membership with trademark transfer. License
integrity follows from trademark control.

\textbf{Foundation governance's backstop has bounds: distribution
remains under steward control.} The Linkerd case of February 2024 is the
structural complement to the NATS case. Linkerd has been CNCF-graduated
since July 2021. Apache 2.0 throughout. Buoyant is the sole commercial
steward and employs every core maintainer, a fact the company
acknowledges publicly. In February 2024, Buoyant announced that the
Linkerd open-source project would no longer publish stable release
artifacts; stable releases would be available only through the
commercial Buoyant Enterprise for Linkerd distribution {[}Buoyant,
2024{]}. The source code remained Apache 2.0. The CNCF charter was not
violated. The trademark backstop did not engage because no relicensing
was attempted. What changed was distribution: the production-grade
artifacts that organisations actually deploy moved from the OSS project
to a paid commercial product.

The mechanism implied here is steward concentration interacting with
venture pressure. When a foundation-graduated project has a single
corporate steward that employs all maintainers, the foundation charter
blocks relicensing but does not block the steward from deciding what the
OSS project publishes versus what the commercial product publishes. When
that steward is venture-backed and faces the same commercial timeline
pressure as single-vendor venture-backed projects in Section 3, the
pressure may find a distribution-side outlet because the licensing-side
outlet is closed. We do not claim this is proven by one event in a
three-project cell; we claim only that the Linkerd case is structurally
consistent with the mechanism, and that the mechanism is invisible if
governance and capital structure are recorded but steward concentration
is not.

The architectural implication: foundation membership alone, without
attention to who employs the maintainers and how that employer is
funded, may be an incomplete signal. The decision instrument of Section
6 is extended to six fields to reflect this.

\textbf{Venture funding is consistent with monetisation pressure on a
misaligned timeline.} Venture funds have five-to-seven-year fund cycles.
Production data infrastructure has multi-decade horizons. The mismatch
resolves in one of two ways. Either the project's commercial steward
generates enough revenue from a managed-service or enterprise tier to
satisfy investor return expectations, or the steward narrows the
open-source surface to protect that revenue against managed-service
competition from cloud providers or other commercial competitors. The
catalogued events are consistent with the second option chosen at scale:
every major SSPL, BSL, or Elastic License adoption between 2018 and 2024
in the catalogue cited cloud-provider competition as the precipitating
cause {[}MongoDB, 2018; Elastic, 2021; HashiCorp, 2023; Redis, 2024{]}.
Buoyant's February 2024 announcement said the same thing in different
terms: the problem was the absence of revenue capture from users running
stable Linkerd in production {[}Buoyant, 2024{]}. Section 7 returns to
the causation-versus-correlation question explicitly.

\textbf{License changes appear to be lagging indicators of revenue
stress.} ArangoDB moved from Apache 2.0 to BSL with v3.12 in early 2024,
citing ``evolving the licensing model for a sustainable future''
{[}ArangoDB, 2024{]}. The framing of the announcement positions the
change as a response to commercial pressure rather than a precipitating
cause. The pattern is similar for CockroachDB's two license moves
(Apache to BSL in 2019, BSL to fully proprietary in 2024) and for
ScyllaDB's December 2024 transition to a source-available licence
{[}ScyllaDB, 2024{]}, each occurring after multi-quarter periods of
strategic repositioning visible in public reporting. We do not claim a
causal direction from this data; the pattern is consistent with license
changes signalling commercial pressure already in progress, rather than
producing it.

\textbf{Forks and reversals add a third dimension.} Three of the largest
2018--2024 license events spawned foundation-backed forks within months:
OpenSearch from Elasticsearch in April 2021, OpenTofu from Terraform in
August 2023, Valkey from Redis in March 2024. Three more produced
reversals. Elastic added AGPLv3 back in August 2024, Redis followed with
AGPLv3 in May 2025, and InfluxDB returned to MIT/Apache 2.0 in January
2025 {[}InfluxData, 2025{]}. The architectural lesson is that adverse
events open mobility paths when the underlying contributor and trademark
structures allow it. Architects who picked components with credible
fork-or-migration paths absorbed events at lower cost than those locked
into proprietary deployments.

\hypertarget{counterfactuals}{%
\section{5. Counterfactuals}\label{counterfactuals}}

Mechanisms gain force from cases that should fail under a naive reading
of the data but do not.

\textbf{PostgreSQL.} Production-relevant since around 2000, license
unchanged since the 1990s. Single-vendor governance is structurally
impossible here because copyright is distributed across thousands of
contributors without CLA aggregation. The PostgreSQL Global Development
Group is a coordinating body, not a copyright holder. Commercial vendors
(EnterpriseDB, Crunchy Data, Supabase, Neon, Tembo) monetise around the
project with no path to relicense it. Distributed copyright is the
structural protection {[}PostgreSQL, 2026{]}.

\textbf{pgvector.} A PostgreSQL extension under MIT, maintained by a
single volunteer developer since 2021. Production-critical for vector
search across many AI deployments. No commercial entity, no venture
backer, no path to commercial relicensing. The protection here is the
absence of monetisation pressure, not the presence of foundation
governance, a different mechanism than PostgreSQL's. The cost is
bus-factor risk: the failure mode would be abandonment, which is a
different shape from license narrowing {[}Kane, 2026{]}.

\textbf{SQLite.} Public domain since 2000. Single-vendor in the strict
sense (Hwaci employs all the developers), but the commercial model sells
warranty-of-title documentation rather than license fees. The protection
is the deliberate choice of a business model that places the software
outside the legal scope of license monetisation {[}SQLite, 2024{]}.

\textbf{Apache Kafka.} Donated to the Apache Software Foundation by
LinkedIn in 2011. Confluent was founded by Kafka's creators but holds no
path to relicense Kafka itself; Confluent's December 2018 license change
applied to peripheral tools it actually owned (Schema Registry, ksqlDB),
not to Kafka {[}Confluent, 2018{]}. The protection is identical in shape
to PostgreSQL's: copyright held by a neutral body, contributions
licensed to the foundation rather than aggregated to a vendor.

\textbf{Caddy.} A web server under Apache 2.0, single-vendor (Light Code
Labs, later Ardan Labs). In 2019 the maintainer attempted to require
commercial licenses for precompiled official binaries. Community
pressure produced a reversal within months and the commercial model
shifted to support services. The 2019 incident is recorded in the
dataset as a reversal entry rather than an adverse event, since the
attempt was withdrawn before taking effect; Caddy is coded as not having
experienced an adverse event in the catalogued window. The case shows
that single-vendor governance combined with active community
accountability can produce the same outcome as foundation governance ---
but the protection is weaker, and depends on the maintainer's
responsiveness {[}Caddy, 2019{]}.

\textbf{The pattern.} Every counterfactual is protected by at least one
of three things: distributed copyright (PostgreSQL, Kafka), absence of
monetisation pressure (pgvector, SQLite, Caddy), active community
accountability (Caddy). None is protected by license type alone. The
structural conditions that protect a project from license events sit
upstream of the license itself. The Linkerd case sharpens this further:
foundation membership without distributed copyright and without dilution
of steward concentration is partial protection, not complete protection.

\hypertarget{decision-instrument}{%
\section{6. Decision Instrument}\label{decision-instrument}}

Architects choosing memory components currently document technical
properties (latency profile, consistency model, query semantics) as
standard procedure. Structural properties are equally measurable but
rarely recorded.

We propose a six-field record per memory component, completed at design
time and updated annually:

\begin{enumerate}
\def\labelenumi{\arabic{enumi}.}
\tightlist
\item
  \textbf{Governance.} Single-vendor, foundation-governed (Apache
  Software Foundation, CNCF, Linux Foundation, or equivalent), or
  multi-vendor.
\item
  \textbf{Capital structure.} Venture-backed, commercial open-source
  company, corporate-sponsored, foundation-funded, or volunteer.
\item
  \textbf{License at adoption.} OSI-approved permissive, OSI-approved
  copyleft, source-available, or proprietary.
\item
  \textbf{Foundation membership.} None, observer or member, incubating,
  or graduated. Foundation membership status correlates with the
  structural backstop described in Section 4.
\item
  \textbf{Fork or migration availability.} None (proprietary or
  single-implementation), thin (source available but no active fork or
  alternative), credible (existing fork or drop-in alternative under
  different governance).
\item
  \textbf{Steward concentration.} Distributed (multiple independent
  corporate or volunteer maintainers, no single party employs more than
  a minority of committers), moderate (one party employs a majority but
  not all core maintainers), or concentrated (one party employs all or
  substantially all core maintainers). The Linkerd case in Section 4
  motivates this field: concentrated steward employment combined with
  venture-backed capital structure can produce adverse outcomes even
  under foundation governance.
\end{enumerate}

The six fields can be summarised in a single-page review template:

\begin{longtable}[]{@{}
  >{\raggedright\arraybackslash}p{(\columnwidth - 6\tabcolsep) * \real{0.2500}}
  >{\raggedright\arraybackslash}p{(\columnwidth - 6\tabcolsep) * \real{0.2500}}
  >{\raggedright\arraybackslash}p{(\columnwidth - 6\tabcolsep) * \real{0.2500}}
  >{\raggedright\arraybackslash}p{(\columnwidth - 6\tabcolsep) * \real{0.2500}}@{}}
\toprule\noalign{}
\begin{minipage}[b]{\linewidth}\raggedright
Field
\end{minipage} & \begin{minipage}[b]{\linewidth}\raggedright
Question
\end{minipage} & \begin{minipage}[b]{\linewidth}\raggedright
Low-risk signal
\end{minipage} & \begin{minipage}[b]{\linewidth}\raggedright
High-risk signal
\end{minipage} \\
\midrule\noalign{}
\endhead
\bottomrule\noalign{}
\endlastfoot
Governance & Who can relicense unilaterally? & foundation / distributed
copyright & single vendor with CLA aggregation \\
Capital structure & What funding pressure exists? & foundation-funded /
volunteer / non-VC corporate & venture-backed \\
License at adoption & What use rights exist today? & OSI permissive or
copyleft & source-available or proprietary \\
Foundation membership & Trademark and legal backstop? & graduated
foundation project & none \\
Fork or migration availability & Can users move if the project changes?
& credible fork or drop-in alternative & none \\
Steward concentration & Who employs the core maintainers? & distributed
across organisations & single company employs the entire core team \\
\end{longtable}

The six fields are observable. Governance is documented in vendor blogs
and contributor agreements. Capital structure is reported by the vendor
or in public filings. License is in the LICENSE file. Foundation
membership is on foundation websites. Fork availability is observable on
GitHub. Steward concentration is observable in the contributor graph and
from the public employment of named maintainers. The cost of recording
these fields per component is small. The cost of not recording them,
when an event arrives, is the engineering time required to discover them
under deadline.

The instrument does not recommend specific components or specific
license types. It records the structural variables that the data shows
correlate with stability. An architect choosing a venture-backed
single-vendor memory component without a credible fork path is accepting
a structural condition that, in the catalogued window, produced an
adverse event for approximately 46 percent of projects in eight years.
An architect choosing a foundation-governed component funded outside the
venture cycle and with distributed steward concentration is accepting a
structural condition that produced events at approximately 2.5 percent
in the same window. An architect choosing a foundation-governed
component with a venture-backed sole steward is accepting an
intermediate condition the cell sample is too small for a stable rate
estimate, but the mechanism described in Section 4 is the architectural
concern. All three choices are defensible. The point is that the choice
should be visible.

For agent memory specifically, the instrument applies most sharply at
the layers where catalogued events have concentrated: graph stores,
vector stores under venture pressure, orchestration frameworks where the
open-source surface intersects with a hosted product, and
service-mesh-like layers where commercial distribution decisions can
shift independently of license. These are the layers where structural
conditions and event rates are highest. They are also where the cost of
an event after deployment is greatest.

A worked example. Consider a May 2026 deployment using PostgreSQL with
pgvector for vector search and Apache Kafka for ingestion. Three
components, three different structural protections: PostgreSQL via
distributed copyright across thousands of contributors, Apache Kafka via
foundation governance and trademark, pgvector via the absence of any
monetisation pressure under a single-volunteer-maintainer model. All
three are permissive-licensed. Two have foundation or
distributed-copyright governance; pgvector carries the structural risk
of bus-factor concentration rather than license-narrowing. The aggregate
risk profile falls in the lower cell of the conditional-rate table, with
bus-factor as a distinct failure mode discussed in Section 5.

Now compare a deployment using a venture-backed single-vendor vector
store, a venture-backed graph store, and a venture-backed orchestration
framework. Three components in the cell that produced events at
approximately 46 percent in the catalogued window. Both deployments may
meet performance and consistency requirements equally well. They are not
equivalent on the structural axis.

\hypertarget{threats-to-validity}{%
\section{7. Threats to Validity}\label{threats-to-validity}}

\textbf{Coding sensitivity analysis.} The most contested coding choice
in the catalogue is the treatment of Linkerd's February 2024
stable-release withdrawal as an adverse event under criterion (b), which
we extended from ``cloud-only product'' to ``commercial product, whether
cloud-hosted or otherwise distributed.'' A reviewer who prefers the
narrower reading would exclude Linkerd from the count. We also report
rates restricted to the narrowest event class --- license changes only,
criterion (a) --- so readers can see how the picture holds up under
maximally conservative event definition.

{\small
\begin{longtable}[]{@{}
  >{\raggedright\arraybackslash}p{(\columnwidth - 14\tabcolsep) * \real{0.1900}}
  >{\raggedright\arraybackslash}p{(\columnwidth - 14\tabcolsep) * \real{0.1000}}
  >{\raggedright\arraybackslash}p{(\columnwidth - 14\tabcolsep) * \real{0.1000}}
  >{\raggedright\arraybackslash}p{(\columnwidth - 14\tabcolsep) * \real{0.1100}}
  >{\raggedright\arraybackslash}p{(\columnwidth - 14\tabcolsep) * \real{0.1100}}
  >{\raggedright\arraybackslash}p{(\columnwidth - 14\tabcolsep) * \real{0.1100}}
  >{\raggedright\arraybackslash}p{(\columnwidth - 14\tabcolsep) * \real{0.1100}}
  >{\raggedright\arraybackslash}p{(\columnwidth - 14\tabcolsep) * \real{0.1100}}@{}}
\toprule\noalign{}
\begin{minipage}[b]{\linewidth}\raggedright
Coding choice
\end{minipage} & \begin{minipage}[b]{\linewidth}\raggedright
Distinct adverse projects
\end{minipage} & \begin{minipage}[b]{\linewidth}\raggedright
Project-level rate
\end{minipage} & \begin{minipage}[b]{\linewidth}\raggedright
Foundation rate
\end{minipage} & \begin{minipage}[b]{\linewidth}\raggedright
Foundation+VC cell
\end{minipage} & \begin{minipage}[b]{\linewidth}\raggedright
Foundation+non-VC cell
\end{minipage} & \begin{minipage}[b]{\linewidth}\raggedright
SV+VC headline
\end{minipage} & \begin{minipage}[b]{\linewidth}\raggedright
Differential
\end{minipage} \\
\midrule\noalign{}
\endhead
\bottomrule\noalign{}
\endlastfoot
Main coding (Linkerd adverse, all event types) & 25/105 & 24\% & 2/43
(4.7\%) & 1/3 & 1/40 (2.5\%) & 23/50 (46\%) & 18.4× \\
Excluding Linkerd & 24/105 & 23\% & 1/43 (2.3\%) & 0/3 & 1/40 (2.5\%) &
23/50 (46\%) & 18.4× \\
License-only events (criterion (a)) & 15/105 & 14\% & 0/43 (0\%) & 0/3 &
0/40 (0\%) & --- & --- \\
\end{longtable}
}

Three things to notice in the table. First, the headline differential of
approximately nineteen-fold (18.4×) between Single-vendor + VC and
Foundation + non-VC cells is \emph{invariant} to the Linkerd coding
choice, because Linkerd sits in the Foundation + VC cell rather than the
Foundation + non-VC cell. A reader who rejects the broader reading of
criterion (b) keeps the same headline. Second, the overall Foundation
governance rate is sensitive to this case. It falls from 4.7 to 2.3
percent without Linkerd, but the Foundation + non-VC sub-cell is
unaffected. Third, restrict to license-only events and zero
foundation-governed projects in the sample experienced an adverse event
at all. The two foundation-governed adverse events in the broader
catalogue are both non-license events: the Hortonworks/Cloudera
ecosystem deprecation (an acquisition outcome) and the Linkerd
distribution change. Foundation governance with trademark transfer
appears to provide complete protection against unilateral relicensing
within this sample. (The license-only row leaves the SV+VC and
differential columns blank because the relevant denominator restriction
would require recomputing the SV+VC cell under the narrower event
definition; the row is included to show the foundation-side result under
maximally conservative event coding, not to recompute the headline
differential.)

\textbf{Sample selection.} The population was constructed from
popularity rankings, foundation membership lists, and a practitioner
landscape, extended by six manual additions documented in Section 2.
Each source has bias. Popularity rankings over-represent commercial
visibility. Foundation lists over-represent projects already
structurally protected. The practitioner landscape is curated. The base
rate of approximately 24 percent should be read as an estimate within
the visible commercial-and-near-commercial open-source ecosystem, not
across all open-source data infrastructure.

\textbf{Survivorship bias.} The intake sources all select for projects
that survived to visibility: popularity rankings, foundation membership
lists, practitioner attention. Projects that died quietly are
systematically under-represented: abandonment without fork, gradual
decline below the inclusion threshold, archive without successor and
without notable press. This bias likely inflates the apparent stability
of the visible ecosystem, particularly for the Volunteer and
Foundation-funded categories where silent decline is more plausible than
headline-grabbing license events. The Volunteer and Foundation-funded
cell rates of zero should be read in that light.

\textbf{Event definition aggregates heterogeneous mechanisms.} The
adverse-event count aggregates license changes, OSS feature removals,
deprecations, acquisitions with narrowing, and repository archivals.
These mechanisms differ in their architectural implications, even though
all five share the property of reducing the operational coverage of the
open-source edition. The license-only column in the sensitivity table
above offers a way to read rates restricted to a single mechanism.
Anyone who finds the broader aggregation unconvincing can anchor on that
column.

\textbf{Capital structure coding at catalogue entry rather than at event
time.} We coded capital structure as of catalogue entry rather than
per-event. For projects without events, coding reflects May 2026 status.
For projects with events occurring years apart (CockroachDB's 2019 and
2024 events are the clearest case), the coding reflects the dominant
capital structure across the period rather than a per-event snapshot. A
more precise rule (capital structure at the time immediately preceding
each event) would refine the venture-funding rate, though we do not
expect the qualitative pattern to shift since most catalogued projects
had stable funding categories across the window. We flag this as a
methodological limitation for any extension of this work.

\textbf{Causation versus correlation.} The conditional rates in Section
3 and the mechanisms in Section 4 establish correlations and articulate
plausible mechanisms. They do not establish causal relationships.
Reverse causation cannot be ruled out from this data: it is possible
that projects which would have undergone adverse events independent of
structural conditions are also the projects that attract venture
funding, rather than venture funding causing the events. The
counterfactuals in Section 5 (PostgreSQL with distributed copyright,
Caddy with community-enforced reversal, NATS with foundation-trademark
backstop) give qualitative support for the structural-mechanism reading
over the reverse-causation reading, but they do not constitute a
controlled test. A causally identified study would require an instrument
independent of project characteristics, which this observational
catalogue does not provide.

\textbf{Event definition.} The boundary between ``license narrowing''
and ``open-core split at launch'' is judgement-dependent. We coded
YugabyteDB's launch with both Apache 2.0 core and BSL enterprise
components as a partial event because the BSL components were always
source-available. We coded Redpanda, DragonflyDB, and Memgraph as
non-events because they were never under permissive licenses in their
open-source form (Memgraph moved from proprietary closed-source to BSL
in 2021, a direction of increased openness rather than decreased). We
coded the OpenObserve transition from Apache 2.0 to AGPL 3.0 in November
2023 as non-adverse because both endpoints are OSI-approved licences
{[}OpenObserve, 2023{]}. Reasonable coders could draw these lines
differently. The qualitative pattern (single-vendor venture governance
correlates with events at much higher rates than foundation governance
with non-venture stewardship) survives most reasonable coding choices,
as the sensitivity table above demonstrates.

\textbf{Time horizon.} Eight years is short for infrastructure with
multi-decade lifetimes. Several catalogued events are still in flux as
of May 2026: the Neo4j AGPL-and-Commons-Clause litigation is on appeal
in the Ninth Circuit with a Free Software Foundation amicus brief filed
in 2025 {[}Neo4j, 2025{]}; the InfluxDB return to permissive licensing
is in public alpha; the IBM stewardship of HashiCorp is post-acquisition
and not yet stable. A ten-year follow-up would refine the rates.

\textbf{Sample size in sub-cells.} Several conditional cells contain
fewer than ten projects (Commercial-OSS at four, Corporate-sponsored at
nine, Foundation-with-VC-steward at three, Volunteer at three). The
rates quoted in those cells should be read as orders of magnitude, not
as precise estimates. The Foundation-with-VC-steward cell observation of
one event in three projects is particularly under-determined; the
architectural argument from that cell rests on the qualitative mechanism
described in Section 4, not on the point estimate.

\textbf{Generalisation to memory specifically.} The 105-project sample
covers data infrastructure broadly, not memory components specifically.
The five 2026 surveys treat memory as a layered system drawing on vector
stores, graph stores, orchestration frameworks, and embedding
infrastructure, all categories represented in our sample. The catalogued
events at the layers most relevant to memory are sufficient to make the
architectural argument, but a memory-only sample would refine the rates.

\hypertarget{open-problems}{%
\section{8. Open Problems}\label{open-problems}}

\textbf{Standardised structural metadata.} The six-field decision
instrument of Section 6 is currently filled by hand.
Software-bill-of-materials standards such as SPDX and CycloneDX address
security provenance and dependency tracking but do not address
governance, capital structure, foundation membership, or steward
concentration. A machine-readable structural-metadata standard, embedded
in component releases, perhaps as an extension of SPDX 3.0 or CycloneDX
1.6, would reduce the cost of architectural review and enable automated
drift detection.

\textbf{Steward-concentration measurement.} The proposed sixth field
(steward concentration) is observable but its operationalisation is not
yet standardised. A reproducible measurement protocol (for example, the
fraction of commits in the trailing twelve months from authors employed
by a single corporate entity) would enable the field to be filled
mechanically rather than by judgement.

\textbf{Longitudinal study.} The catalogued window (2018--May 2026)
covers the rise of cloud managed-service competition as a license-event
driver. The next eight years will test whether foundation backstops,
fork mobility, reversal pathways, and steward-diversification efforts
(such as the post-event Linkerd governance review by the CNCF Technical
Oversight Committee) change the equilibrium. A repeat study in 2034
would establish whether the increase in annualised incidence between
2018--2020 and 2024--May 2026 continues, plateaus, or reverses.

\textbf{Integration with security primitives.} Lin et al.'s {[}Lin et
al., 2026{]} nine governance primitives for memory security --- memory
unit abstraction, write gate, provenance metadata, versioning, trust
labels, principal scoping, rollback, deletion semantics,
internal-channel observability --- operate at runtime. The structural
variables documented here operate at the supply chain. The two compose:
which primitives are implementable in a given deployment depends on the
structural status of its components. A combined architectural-review
checklist covering both layers would be a useful next artifact.

\textbf{Memory-specific event catalogue.} This paper's sample is data
infrastructure broadly. A focused catalogue of license and
sustainability events affecting memory components specifically (vector
stores, graph stores, embedding infrastructure, orchestration,
service-mesh-adjacent layers) would refine rates for the use case the
2026 surveys describe.

The five 2026 memory surveys each treat their subject thoroughly within
the scope they set for themselves. That scope does not include the
structural variables that, on this data, correlate most strongly with
whether a memory component will still be a viable production choice
eighteen months from now. Adding governance, capital structure, license,
foundation membership, fork availability, and steward concentration to
architectural review costs little and improves a decision the surveys
leave unaddressed.

\begin{center}\rule{0.5\linewidth}{0.5pt}\end{center}

\hypertarget{references}{%
\section{References}\label{references}}

{[}Anonymous, 2026{]} \emph{LLM Agent Memory: A Survey from a Unified
Representation.} OpenReview, 2026.
https://openreview.net/forum?id=KPs1EgGKcT

{[}ArangoDB, 2024{]} ArangoDB. \emph{Evolving ArangoDB's Licensing Model
for a Sustainable Future.} February 2024.
https://arango.ai/blog/update-evolving-arangodbs-licensing-model-for-a-sustainable-future/

{[}Buoyant, 2024{]} Buoyant. \emph{Announcing Linkerd 2.15 with mesh
expansion, native sidecars, and SPIFFE.} 21 February 2024.
https://linkerd.io/2024/02/21/announcing-linkerd-2.15/ See also the
eight-month retrospective: \emph{Towards a Sustainable Service Mesh.} 23
October 2024. https://linkerd.io/2024/10/23/making-linkerd-sustainable/

{[}Caddy, 2019{]} Caddy Community. \emph{Caddy License for Commercial
Use.} GitHub Issue \#2786 and community thread, October 2019.
https://caddy.community/t/caddy-license-for-commercial-use/17170

{[}CNCF, 2025{]} Cloud Native Computing Foundation. \emph{Protecting
NATS and the Integrity of Open Source: CNCF's Commitment to the
Community.} 1 May 2025.
https://www.cncf.io/blog/2025/05/01/protecting-nats-and-the-integrity-of-open-source-cncfs-commitment-to-the-community/

{[}CNCF, 2026{]} Cloud Native Computing Foundation. \emph{Graduated and
Incubating Projects.} https://www.cncf.io/projects/

{[}Cockroach Labs, 2024{]} Cockroach Labs. \emph{CockroachDB Software
License.} Announced 15 August 2024 (effective with v24.3 in November
2024). The 2024 announcement blog post is no longer available at its
original URL; the license terms and migration timeline are documented in
Oxide RFD 0508. https://rfd.shared.oxide.computer/rfd/0508

{[}Confluent, 2018{]} Confluent. \emph{License Changes for Confluent
Platform.} 14 December 2018.
https://www.confluent.io/blog/license-changes-confluent-platform/

{[}DB-Engines, 2026{]} DB-Engines. \emph{DB-Engines Ranking, May 2026.}
https://db-engines.com/en/ranking

{[}Du et al., 2026{]} Y. Du et al.~\emph{Memory for Autonomous LLM
Agents: Mechanisms, Evaluations, and Open Problems.} arXiv:2603.07670,
March 2026.

{[}Eghbal, 2020{]} N. Eghbal. \emph{Working in Public: The Making and
Maintenance of Open Source Software.} Stripe Press, 2020.

{[}Elastic, 2021{]} Elastic. \emph{Licensing Change.} 14 January 2021.
https://www.elastic.co/blog/licensing-change

{[}Elastic, 2024{]} Elastic. \emph{Elasticsearch is Open Source, Again.}
29 August 2024.
https://www.elastic.co/blog/elasticsearch-is-open-source-again

{[}HashiCorp, 2023{]} HashiCorp. \emph{License FAQ.} August 2023.
https://www.hashicorp.com/en/license-faq

{[}IBM, 2025{]} IBM. \emph{IBM Completes Acquisition of HashiCorp.} 27
February 2025.
https://newsroom.ibm.com/2025-02-27-ibm-completes-acquisition-of-hashicorp

{[}InfluxData, 2025{]} InfluxData. \emph{InfluxDB 3 Open Source Now in
Public Alpha Under MIT/Apache 2 License.} January 2025.
https://community.influxdata.com/t/influxdb-3-open-source-now-in-public-alpha-under-mit-apache-2-license/55208

{[}Kane, 2026{]} A. Kane. \emph{pgvector LICENSE.}
https://github.com/pgvector/pgvector/blob/master/LICENSE

{[}Lin et al., 2026{]} H. Lin et al.~\emph{Toward Mnemonic Sovereignty:
A Survey on the Security of Long-Term Memory in LLM Agents.}
arXiv:2604.16548, April 2026.

{[}Linux Foundation, 2024{]} Linux Foundation. \emph{Linux Foundation
Launches Open Source Valkey Community.} March 2024.
https://www.linuxfoundation.org/press/linux-foundation-launches-open-source-valkey-community

{[}Luo et al., 2026{]} Y. Luo et al.~\emph{From Storage to Experience.}
ICLR 2026 MemAgents Workshop. https://openreview.net/forum?id=l9Ly41xxPb

{[}Mem0, 2026{]} Mem0. \emph{Open Source: Migrating to the New Memory
Algorithm (v2 to v3).} April 2026.
https://docs.mem0.ai/migration/oss-v2-to-v3

{[}MongoDB, 2018{]} MongoDB. \emph{MongoDB Issues New Server Side Public
License for MongoDB Community Server.} 16 October 2018.
https://www.mongodb.com/company/newsroom/press-releases/mongodb-issues-new-server-side-public-license-for-mongodb-community-server

{[}Neo4j, 2025{]} T. Claburn. \emph{Free Software Foundation Defends
AGPLv3 in Neo4j Appeal.} The Register, 4 March 2025.
https://www.theregister.com/2025/03/04/free\_software\_foundation\_agplv3/

{[}OpenObserve, 2023{]} OpenObserve. \emph{What are Apache, GPL and AGPL
licenses and why OpenObserve moved from Apache to AGPL.} November 2023.
https://openobserve.ai/blog/what-are-apache-gpl-and-agpl-licenses-and-why-openobserve-moved-from-apache-to-agpl/

{[}OpenTofu, 2023{]} OpenTofu. \emph{The OpenTofu Fork Is Now
Available.} September 2023.
https://opentofu.org/blog/the-opentofu-fork-is-now-available/

{[}Ostrom, 1990{]} E. Ostrom. \emph{Governing the Commons: The Evolution
of Institutions for Collective Action.} Cambridge University Press,
1990.

{[}PostgreSQL, 2026{]} PostgreSQL Global Development Group.
\emph{About.} https://www.postgresql.org/about/

{[}Pracdata, 2024{]} \emph{Open Source Data Engineering Landscape 2024.}
https://www.pracdata.io/p/open-source-data-engineering-landscape-2024

{[}Redis, 2024{]} Redis. \emph{Redis Adopts Dual Source-Available
Licensing.} 20 March 2024.
https://redis.io/blog/redis-adopts-dual-source-available-licensing/

{[}Redis, 2025{]} Redis. \emph{Redis is Now Available under the AGPLv3
Open Source License.} 1 May 2025. https://redis.io/blog/agplv3/

{[}Schweik \& English, 2012{]} C.M. Schweik and R.C. English.
\emph{Internet Success: A Study of Open-Source Software Commons.} MIT
Press, 2012.

{[}ScyllaDB, 2024{]} ScyllaDB. \emph{Why We're Moving to a Source
Available License.} 18 December 2024.
https://www.scylladb.com/2024/12/18/why-were-moving-to-a-source-available-license/

{[}Sentry, 2023{]} Sentry. \emph{Introducing the Functional Source
License: Freedom Without Free-Riding.} 17 November 2023.
https://blog.sentry.io/introducing-the-functional-source-license-freedom-without-free-riding/

{[}SQLite, 2024{]} SQLite. \emph{Copyright Notice.}
https://sqlite.org/copyright.html

{[}TechCrunch, 2023{]} K. Wiggers. \emph{ScyllaDB raises \$43M to scale
its NoSQL database platform.} TechCrunch, 17 October 2023.
https://techcrunch.com/2023/10/17/scylladb-raises-43m-to-scale-its-nosql-database-platform/
Cited for cumulative venture funding (\textasciitilde\$103M as of
October 2023) used in capital-structure coding of ScyllaDB.

{[}Wu, 2026{]} J. Wu. \emph{Memory in the LLM Era: Modular Architectures
and Abstractions.} arXiv:2604.01707, April 2026.

{[}Yin et al., 2022{]} L. Yin, M. Chakraborti, Y. Yan, C. Schweik, S.
Frey, and V. Filkov. \emph{Open Source Software Sustainability:
Combining Institutional Analysis and Socio-Technical Networks.} Proc.
ACM Hum.-Comput. Interact. 6, CSCW2, Article 404 (November 2022).
https://doi.org/10.1145/3555129

{[}Zep, 2025{]} Zep. \emph{Announcing a New Direction for Zep's Open
Source Strategy.} 2 April 2025.
https://blog.getzep.com/announcing-a-new-direction-for-zeps-open-source-strategy/

\end{document}